\documentclass[12pt]{article}
 \usepackage{epsfig}
 \def\be{\begin{equation}}
 \def\ee{\end{equation}}
 \def\bea{\begin{eqnarray}}
 \def\eea{\end{eqnarray}}
 \usepackage{graphicx}

 \catcode`\@=11
 \def\lsim{\mathrel{\mathpalette\@versim<}}
 \def\gsim{\mathrel{\mathpalette\@versim>}}
 \def\@versim#1#2{\vcenter{\offinterlineskip
 \ialign{$\m@th#1\hfil##\hfil$\crcr#2\crcr\sim\crcr } }}
 \catcode`\@=12

 \catcode`@=12
\begin{document}
 \thispagestyle{empty}
 \begin{flushright}
 UCRHEP-T587\\
 Jan 2018\
 \end{flushright}
 \vspace{0.4in}
 \begin{center}
 {\LARGE \bf Alternative $[SU(3)]^4$ Model of\\ Leptonic Color 
 and Dark Matter\\}
 \vspace{1.0in}
 {\bf Corey Kownacki, Ernest Ma, Nicholas Pollard, Oleg Popov, and 
Mohammadreza Zakeri\\}
 \vspace{0.2in}
 {\sl Physics and Astronomy Department,\\ 
 University of California, Riverside, California 92521, USA\\}
 \end{center}
 \vspace{1.0in}

\begin{abstract}\
The alternative $[SU(3)]^4$ model of leptonic color and dark matter 
is discussed.  It unifies at $M_U \sim 10^{14}$ GeV and has the low-energy 
subgroup 
$SU(3)_q \times SU(2)_l \times SU(2)_L \times SU(2)_R \times U(1)_X$ 
with $(u,h)_R$ instead of $(u,d)_R$ as doublets under $SU(2)_R$.  It has the 
built-in global $U(1)$ dark symmetry which is generalized $B-L$. 
In analogy to $SU(3)_q$ quark triplets, it has $SU(2)_l$ hemion doublets 
which have half-integral charges and are confined by $SU(2)_l$ gauge 
bosons (stickons).  In analogy to quarkonia, their vector bound states 
(hemionia) are uniquely suited for exploration at a future $e^-e^+$ collider.  
\end{abstract}

 \newpage
 \baselineskip 24pt
\section{Introduction}
To venture beyond the Standard Model (SM) of quarks and leptons, there have 
been many trailblazing ideas.  One is the notion of grand unification, i.e. 
the embedding of the SM gauge symmetry $SU(3)_C \times SU(2)_L \times U(1)_Y$ 
in a single larger symmetry such as $SU(5) \sim E_4$, $SO(10) \sim E_5$, or 
$E_6$.  There are indeed very many papers devoted to this topic.  Less 
visited are the symmetries $[SU(3)]^N$, where $N=3,4,6$ have been 
considered~\cite{dgg84,bhp86,mmz04,bmw04,kmppz17,kmppz18,m05,m04}. 
Another idea is that the $SU(2)_R$ quark doublet may not be $(u,d)_R$ but 
rather $(u,h)_R$ where $h$ is an exotic quark of charge $-1/3$.  This was 
originally motivated by superstring-inspired $E_6$ models~\cite{m87,bhm87} 
and later generalized to nonsupersymmetric 
models~\cite{klm09,klm10,bmw14,kmppz17-1}, 
but is easily implemented in $[SU(3)]^N$ models.  A third idea is quark-lepton 
interchange symmetry~\cite{fl90,flv91} which assumes $SU(3)_l$ for leptons 
in parallel to $SU(3)_q$ for quarks, but with $SU(3)_l$ broken to 
$SU(2)_l \times U(1)_{Y_l}$.  This is naturally embedded in 
$[SU(3)]^4$~\cite{bmw04} and implies that only one component of the color 
lepton triplet is free, i.e. the observed lepton, whereas the other two color 
components (with half-integral charges) are confined in analogy to the three 
color components of a quark triplet.  Finally a fourth idea has been put 
forward recently~\cite{kmppz18,dhqvv18} that a dark symmetry may 
exist within $[SU(3)]^N$ itself or perhaps $[SU(3)]^N \times U(1)$.
This new insight points to the 
possible intrinsic unity of matter with dark matter~\cite{b12,m13,m18}.

In this paper, all four of the above ideas are incorporated into a single 
consistent framework based on the symmetry 
$SU(3)_q \times SU(3)_L \times SU(3)_l \times SU(3)_R$.  The three families 
of quarks and leptons are contained in the bifundamental chain 
$(3,3^*,1,1)+(1,3,3^*,1)+(1,1,3,3^*)+(3^*,1,1,3)$ which also include other 
fermions beyond the SM.  This unifying symmetry is broken by two bifundamental 
scalars at $M_U$ to 
$SU(3)_q \times SU(2)_l \times SU(2)_L \times SU(2)_R \times U(1)_X$ 
in such a way that a residual global $U(1)_D$ symmetry remains.  This 
important property guarantees that a dark sector exists for a set of 
fermions, scalars, and vector gauge bosons.  Because of the necessary 
particle content of $[SU(3)]^4$, this $U(1)_D$ may be identified as 
generalized $B-L$~\cite{m15}, under which quarks have charge 1/3 and leptons 
have charge $-1$, but the other particles have different values. 

At $M_R$ of order a TeV, $SU(2)_R \times U(1)_X$ 
is broken to $U(1)_Y$ of the SM, with particle content of the SM plus possible 
light particles transforming under the leptonic color $SU(2)_l$ symmetry. 
We will discuss their impact on cosmology as well as their possible 
revelation at a future $e^-e^+$ collider, following closely our previous 
work~\cite{kmppz17} on the subject.  We will also consider the 
phenomenology associated with the $SU(2)_R$ gauge symmetry and the possible 
dark-matter candidates of this model.

\section{Fermion Content and Dark Symmetry}
All fermions belong to bitriplet representations $(3,3^*)$ 
under $SU(3)_A \times SU(3)_B$, where $SU(3)_A$ acts vertically from up to 
down with $I_{3A} = (1/2,-1/2,0)$ and 
$Y_A = (1,1,-2)/(2\sqrt{3})$, 
and $SU(3)_B$ horizontally from left to right with 
$I_{3B} = (-1/2,1/2,0)$ and $Y_B$ = $(-1,-1,2)/(2\sqrt{3})$. 
The dark symmetry we will consider is 
\begin{equation}
D = \sqrt{3} (-2 Y_L + \sqrt{3} I_{3R} + Y_R - 2Y_l).
\end{equation}
Under $SU(3)_q \times SU(3)_L \times SU(3)_l \times SU(3)_R$, the fermion 
content of our model is then given by
\begin{eqnarray}
&& q \sim (3,3^*,1,1) \sim \pmatrix{d & u & h \cr d & u & h \cr d & u & h}, 
~ ~ D_q \sim \pmatrix{1 & 1 & -2 \cr 1 & 1 & -2 \cr 1 & 1 & -2}, \\ 
&& l \sim (1,3,3^*,1) \sim \pmatrix{x_1 & x_2 & \nu \cr y_1 & y_2 & e \cr z_1 
& z_2 & n}, ~ ~ D_{l} \sim \pmatrix{0 & 0 & -3 \cr 0 & 0 & -3 \cr 3 & 3 & 0}, \\ 
&& l^c \sim (1,1,3,3^*) \sim \pmatrix{z_1^c & y_1^c & x_1^c \cr z_2^c & y_2^c 
& x_2^c \cr n^c & e^c & \nu^c}, ~ ~  
D_{l^c} \sim \pmatrix{-3 & 0 & 0 \cr -3 & 0 & 0 \cr 0 & 3 & 3}, \\ 
&& q^c \sim (3^*,1,1,3) \sim \pmatrix{h^c & h^c & h^c \cr u^c & u^c & u^c \cr 
d^c & d^c & d^c}, ~ ~ 
D_{q^c} \sim \pmatrix{2 & 2 & 2 \cr -1 & -1 & -1 \cr -1 & -1 & -1}, 
\end{eqnarray}
where $u$ has charge $2/3$, $d,h$ have charge $-1/3$, $x,z$ have charge 1/2, 
$y$ has charge $-1/2$, $\nu,n$ have charge 0, and $e$ has charge $-1$.
Using
\begin{equation}
R_D = (-1)^{D+2j},
\end{equation}
we see that $u,u^c,d,d^c,\nu,\nu^c,e,e^c,z,z^c$ are even, and 
$h,h^c,x,x^c,y,y^c,n,n^c$ are odd.  Further, the gauge bosons which take $h$ 
to $u,d$ in $SU(3)_L$ and $h^c$ to $u^c,d^c$ in $SU(3)_R$ are odd, 
as well as the corresponding ones in $SU(3)_l$, and the others even, including 
all those of the SM.  Hence $R_D$ would remain a good symmetry for dark 
matter provided that the scalar sector responsible for the symmetry 
breaking obeys it as well.

The scalar bitriplets responsible for the masses of the fermions in Eqs.~(2) 
to (5) come from three chains, each of the form 
$(3,1,3^*,1)+(1,3,1,3^*)+(3^*,1,3,1)+(1,3^*,1,3)$.  Specifically,
\begin{eqnarray}
&& \phi^{(1,3,5)} \sim (1,3,1,3^*) \sim \pmatrix{\eta^0 & \phi_2^+ & 
\phi_1^0 \cr \eta^- & \phi_2^0 & \phi_1^- \cr \chi^0 & \chi^+ & \lambda^0}, 
~~~ D_\phi \sim \pmatrix{-3 & 0 & 0 \cr -3 & 0 & 0 \cr 0 & 3 & 3}, \\ 
&& \bar{\phi}^{(2,4,6)} \sim (1,3^*,1,3) \sim \pmatrix{\bar{\eta}^0 & \eta^+ & 
\bar{\chi}^0 \cr \phi_2^- & \bar{\phi}_2^0 & \chi^- \cr \bar{\phi}_1^0 & 
\phi_1^+ & \bar{\lambda}^0}, ~~~ D_{\bar{\phi}} \sim \pmatrix{3 & 3 & 0 \cr 
0 & 0 & -3 \cr 0 & 0 & -3}. 
\end{eqnarray}
From the $q^c q \phi$ terms, we obtain masses of $hh^c$ from 
$\langle \chi^0 \rangle^{(1)}$, $dd^c$ from $\langle \phi_1^0 \rangle^{(3)}$, 
$uu^c$ from $\langle \phi_2^0 \rangle^{(5)}$.  From the $l l^c \bar{\phi}$ 
terms, we obtain masses of $nn^c,zz^c$ from 
$\langle \bar{\chi}^0 \rangle^{(2)}$, $\nu \nu^c, xx^c$ from 
$\langle \bar{\phi}_1^0 \rangle^{(4)}$, $ee^c, yy^c$ from 
$\langle \bar{\phi}_2^0 \rangle^{(6)}$.  It is clear that $D$ and thus 
$R_D$ remain unbroken by the above vacuum expectation values.

\section{Symmetry Breaking Pattern}
We consider the breaking of $[SU(3)]^4$ at $M_U$ by two scalar bitriplets, 
one transforming as $\phi^{L+} \sim (1,3,3^*,1) \sim l$, belonging to a 
chain in parallel to the fermions, the other transforming as 
$\phi^{R-} \sim (1,1,3,3^*) \sim l^c$, belonging to a chain with an additional 
overall imposed assignment of odd $R_D$, i.e. an additional $Z_2$ 
factor~\cite{kmppz18}.  This preserves the relative $R_D$ 
among its components, but prevents it from coupling to the fermions. Using 
$\langle \phi^{L+}_{33} \rangle$ with even $R_D$ to break 
$SU(3)_L \times SU(3)_l$ to 
$SU(2)_L \times SU(2)_l \times U(1)_{(Y_L+Y_l)/\sqrt{2}}$ and 
$\langle \phi^{R-}_{33} \rangle$ which also has even $R_D$ to break 
$SU(3)_l \times SU(3)_R$ to 
$SU(2)_l \times SU(2)_R \times U(1)_{(Y_l+Y_R)/\sqrt{2}}$, the resulting 
theory preserves $R_D$.  Assuming also that all the particles of the 
chain associated with $\phi^{R-}$ are superheavy, the low-energy theory 
with the residual gauge symmetry 
$SU(3)_q \times SU(2)_l \times SU(2)_L \times SU(2)_R \times U(1)_X$, 
where $X = (Y_L+Y_R+Y_l)/\sqrt{3}$, also preserves $D$.

Since there are three fermion chains, and five scalar chains, the $b$ 
coefficients for the renormalization-group running of each $SU(3)$ 
gauge coupling are all given by
\begin{equation}
b = -11 + {2 \over 3} \left( {1 \over 2} \right) (2)(3)(3) + 
{1 \over 3} \left( {1 \over 2} \right) (2) (3) (5) = 0.
\end{equation} 
This shows that we have a possible finite field theory~\cite{mmz04} 
above $M_U$.

At $M_R$, the $SU(2)_R \times U(1)_X$ gauge symmetry is broken to 
$U(1)_Y$ of the SM, where $Y = I_{3R} - X$, by an $SU(2)_R$ doublet whose 
neutral component is a linear combination of $\chi^0$ from $\phi^{(1)}$, 
the conjugate of $\bar{\chi}^0$ from $\bar{\phi}^{(2)}$, and 
$\phi^{R+}_{31}$ from the $(1,1,3,3^*)$ component of the chain containing 
$\phi^{L+}$ discussed previously.  From the allowed antisymmetric trilinear 
term $l^c l^c \phi^{R+}$, the mass term $x_1^c y_2^c - x_2^c y_1^c$ is 
then obtained.  Note that the correponding mass term $x_1 y_2 - x_2 y_1$ 
is superheavy because it comes from $\langle \phi^{L+}_{33} \rangle$.
Note also that the corresponding term $l^c l^c \phi^{R-}$ is forbidden 
because of the overall assignment of odd $R_D$ for $\phi^{R-}$. 
Finally the symmetry $SU(2)_L \times U(1)_Y$ is broken by two $SU(2)_L$ 
doublets to $U(1)_{em}$ with $Q = I_{3L} + Y$. 

\section{Renormalization-Group Running of Gauge Couplings}
The renormalization-group evolution of the gauge couplings is dictated at 
leading order by
\begin{equation}
{1 \over \alpha_i(\mu)} = {1 \over \alpha_i(\mu')} + {b_i \over 2\pi} 
\ln \left( {\mu' \over \mu} \right),
\end{equation}
where $b_i$ are the one-loop beta-function coefficients. 
From $M_U$ to $M_R$, we assume that all fermions are light except the three 
families of $(x,y)$ hemions.  As for the scalars, we assume that only the 
following multiplets are light under $SU(2)_L \times SU(2)_R \times U(1)_X$: 
1 copy of $(1,2,-1/2)$, 6 copies of $(2,2,0)$, 3 copies of $(2,1,-1/2)$, 
and 4 copies of $(2,1,1/2)$.  This choice requires fine tuning in the scalar 
sector as in other models of grand unification such as $SU(5)$ and $SO(10)$. 
As a result, the five $b$ coefficients are given by
\begin{eqnarray}
b_q &=& -11 + {2 \over 3} \left( {1 \over 2} \right) (6)(3) = -5, \\ 
b_l &=& -{22 \over 3} + {2 \over 3} \left( {1 \over 2} \right) (4)(3) = 
-{10 \over 3}, \\ 
b_L &=& -{22 \over 3} + {2 \over 3} \left( {1 \over 2} \right) (3+1)(3) 
+ {1 \over 3} \left( {1 \over 2} \right) [7 + 6(2)] = -{1 \over 6}, \\ 
b_R &=& -{22 \over 3} + {2 \over 3} \left( {1 \over 2} \right) (3+2+1)(3) 
+ {1 \over 3} \left( {1 \over 2} \right) [1+6(2)] = {5 \over 6}, \\ 
b_X &=& {2 \over 3} \left[ {1 \over 6} (3) + {1 \over 6} (3) + {1 \over 4} (4) 
+ {1 \over 4} (4) \right] (3) + {1 \over 3} \left( {1 \over 4} \right) 
[2+7(2)] = {22 \over 3}.
\end{eqnarray} 

From $M_R$ to $M_Z$, we assume the SM quark and lepton content together with 
1 copy of $(x^c,y^c)$ hemions and two $SU(2)_L$ Higgs scalar doublets.  The 
massless $SU(2)_l$ stickons are of course included but they affect only 
$\alpha_l$.  The four $b$ coefficients are then
\begin{eqnarray}
b_q &=& -11 + {2 \over 3} \left( {1 \over 2} \right) (4)(3) = -7, \\ 
b_l &=& -{22 \over 3} + {2 \over 3} \left( {1 \over 2} \right) (2) = 
-{20 \over 3}, \\ 
b_L &=& -{22 \over 3} + {2 \over 3} \left( {1 \over 2} \right) (3+1)(3) + 
{1 \over 3} \left( {1 \over 2} \right) (2) = -3, \\ 
b_Y &=& {1 \over 2} \left[ {2 \over 3} \left\{ {10 \over 3}(3) + 
{1 \over 4}(4) \right\} + {1 \over 3} \left( {1 \over 4} \right)(4) \right] = 
{23 \over 6},
\end{eqnarray}
where a factor of 1/2 has been inserted to normalize $b_Y$.  The boundary 
condition at $M_R$ for $SU(2)_R \times U(1)_X$ to become $U(1)_Y$ is
\begin{equation}
{2 \over \alpha_Y(M_R)} = {1 \over \alpha_R(M_R)} + {1 \over \alpha_X(M_R)}.
\end{equation}
We then obtain
\begin{eqnarray}
&& {1 \over \alpha_q(M_Z)} = {1 \over \alpha_U} - {7 \over 2 \pi} 
\ln {M_R \over M_Z} - {5 \over 2 \pi} \ln {M_U \over M_R}, \\ 
&& {1 \over \alpha_L(M_Z)} = {1 \over \alpha_U} - {3 \over 2 \pi} 
\ln {M_R \over M_Z} - {1 \over 6(2 \pi)} \ln {M_U \over M_R}, \\
&& {1 \over \alpha_Y(M_Z)} = {1 \over \alpha_U} +{23 \over 6(2\pi)} 
\ln {M_R \over M_Z} + {49 \over 12(2\pi)} \ln {M_U \over M_R}.
\end{eqnarray}

\begin{figure}[htb]
\vspace*{-2cm}
\hspace*{-2cm}
\includegraphics[scale=1]{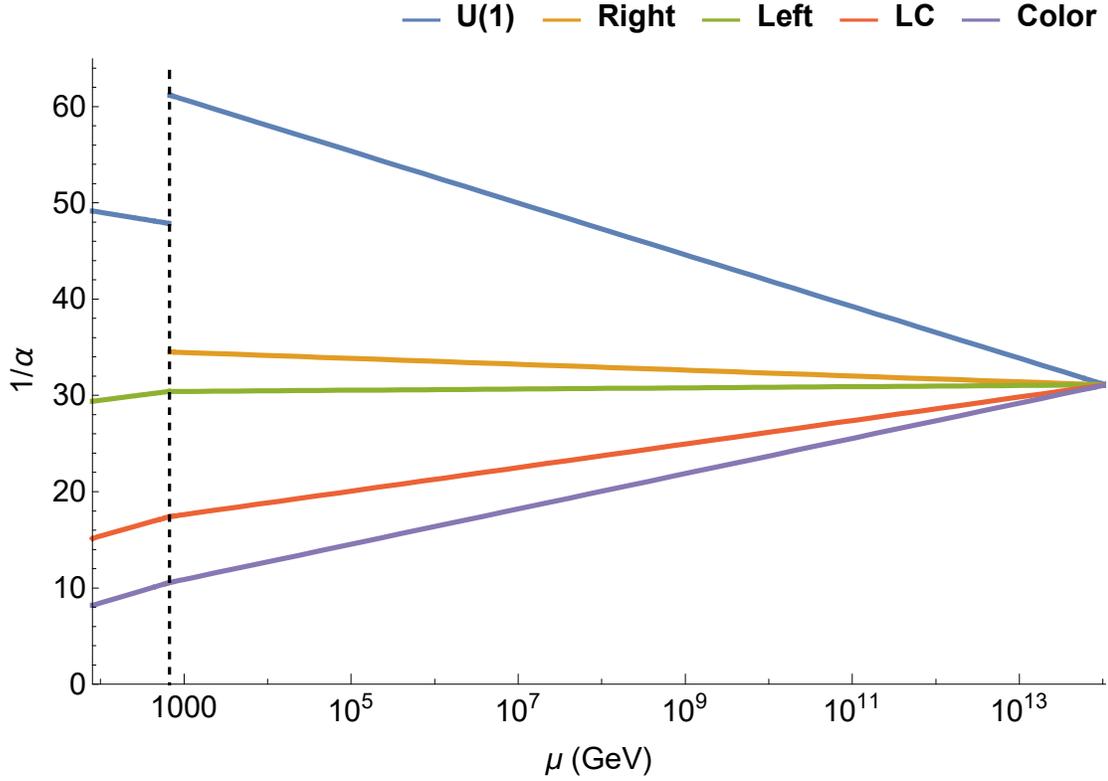}
\vspace*{-15cm}
\caption{Evolution of $\alpha_i^{-1}$ as a function of energy scale.}
\end{figure}
Using the experimental inputs
\begin{eqnarray}
\alpha_q(M_Z) &=& 0.1185, \\ 
\alpha_L(M_Z) &=& (\sqrt{2}/\pi)G_F M_W^2 = 0.0339, \\ 
\alpha_Y(M_Z) &=& 2\alpha_L(M_Z)\tan^2 \theta_W = 0.0204,
\end{eqnarray}
where a factor of 2 has been used to normalize $\alpha_Y$, we find
\begin{eqnarray}
{1 \over 0.0339} - {1 \over 0.1185} = 21.06 &=& {4 \over 2 \pi} 
\ln {M_R \over M_Z} + {29 \over 6(2 \pi)} \ln {M_U \over M_R}, \\ 
{1 \over 0.0204} - {1 \over 0.0339} = 19.52 &=& {41 \over 6(2 \pi)} 
\ln {M_R \over M_Z} + {17 \over 4(2 \pi)} \ln {M_U \over M_R}.
\end{eqnarray}
This implies $M_R \simeq 600$ GeV and $M_U \simeq 10^{14}$ GeV, as shown in 
Fig.~1.  
The 5 lines emanating from a common point at $10^{14}$ GeV represent 
$U(1)_X$, $SU(2)_R$, $SU(2)_L$, $SU(2)_l$, and $SU(3)_q$ from top to bottom. 
The line between $M_R$ and $M_Z$ represents normalized $U(1)_Y$. 
Since there are uncertainties (both theoretical and experimental) in  
the above estimate, the value of $M_R$ should not be taken too literally, 
but rather an indication that particles transforming under $SU(2)_R$ have 
masses of an order of magnitude greater than those of the SM. 
As a result, $\alpha_U = 0.0322$.  Using
\begin{equation}
{1 \over \alpha_R(M_R)} = {1 \over \alpha_U} + 
{5 \over 6(2\pi)} \ln {M_U \over M_R},
\end{equation}
we obtain $\alpha_R(M_R) = 0.0290$. Using
\begin{equation}
{1 \over \alpha_l(M_Z)} - {1 \over \alpha_q(M_Z)} = 
{1 \over 3(2\pi)} \ln {M_R \over M_Z} + {5 \over 3(2\pi)} \ln {M_U \over M_R},
\end{equation}
we obtain $\alpha_l = 0.0650$, implying a confining scale of about 
0.4 MeV from leptonic color.  This is significantly different from the 
result of the $[SU(3)]^4$ 
model with $M_R=M_U$, where it is a few keV~\cite{bmw04,kmppz17}.

\section{Low-Energy Particle Content}
The particles of this model at or below a few TeV are listed in Table 1 under 
$SU(3)_q \times SU(2)_l \times SU(2)_L \times SU(2)_R \times U(1)_X \times D$, 
where $X=(Y_L+Y_R+Y_l)/\sqrt{3}$ (each $Y$ normalized according to 
$\sum Y^2 = 1/2$), $D=\sqrt{3}(-2Y_L+\sqrt{3}I_{3R}+Y_R-2Y_l)$, 
and $Q = I_{3L} + I_{3R} - X$. 
\begin{table}[htb]
\caption{Particle content of proposed model.}
\begin{center}
\begin{tabular}{|c|c|c|c|c|c|c||c|c|}
\hline
particles & $SU(3)_q$ & $SU(2)_l$ & $SU(2)_L$ & $SU(2)_R$ & $U(1)_X$ 
& $D$ & $S$ & $I_{3R} + S$ \\
\hline
$(u,d)_L$ & 3 & 1 & 2 & 1 & $-1/6$ & (1,1) & 1/3 & 1/3 \\
$(u,h)_R$ & $3$ & 1 & 1 & 2 & $-1/6$ & $(1,-2)$ & $-1/6$ & $(1/3,-2/3)$ \\
$d_R$ & $3$ & 1 & 1 & 1 & $1/3$ & 1 & 1/3 & 1/3 \\
$h_L$ & 3 & 1 & 1 & 1 & $1/3$ & $-2$ & $-2/3$ & $-2/3$ \\ 
\hline
$(\nu,l)_L$ & 1 & 1 & 2 & 1 & $1/2$ & $(-3,-3)$ & $-1$ & $-1$ \\
$(n,l)_R$ & 1 & 1 & 1 & 2 & $1/2$ & $(0,-3)$ & $-1/2$ & $(0,-1)$ \\ 
$\nu_R$ & 1 & 1 & 1 & 1 & $0$ & $-3$ & $-1$ & $-1$ \\
$n_L$ & 1 & 1 & 1 & 1 & 0 & $0$ & 0 & 0 \\
\hline
$(z,y)_R$ & 1 & 2 & 1 & 2 & 0 & $(3,0)$ & 1/2 & $(1,0)$ \\ 
$x_R$ & 1 & 2 & 1 & 1 & $-1/2$ & 0 & $0$ & $0$ \\ 
$z_L$ & 1 & 2 & 1 & 1 & $-1/2$ & $3$ & 1 & 1 \\ 
\hline
$(\phi_1^0,\phi_1^-)$ & 1 & 1 & 2 & 1 & $1/2$ & $0$ & 0 & 0 \\
$(\chi^+,\chi^0)$ & 1 & 1 & 1 & 2 & $-1/2$ & $(3,0)$ & 1/2 & (1,0) \\ 
$(\eta,\Phi_2)$ & 1 & 1 & 2 & 2 & $0$ & $(-3,0)$ & $-1/2$ & $(-1,0)$ \\
$\lambda^0$ & 1 & 1 & 1 & 1 & $0$ & $3$ & 1 & 1 \\
\hline
\end{tabular}
\end{center}
\end{table}
The $SU(2)_L \times SU(2)_R$ scalar bidoublet contains the $SU(2)_L$ 
doublets $\eta = (\eta^0,\eta^-)$ and $\Phi_2 = (\phi_2^+,\phi_2^0)$, with 
$\eta$ heavy at the $M_R$ scale.  Because of the assumed symmetry breaking 
pattern, our model actually possesses a conserved global symmetry
\begin{equation}
S = {1 \over \sqrt{3}} (Y_R - 2Y_L - 2Y_l)
\end{equation}
before $SU(2)_R$ breaking, even though the corresponding gauge symmetry 
has been broken.  Whereas both $S$ and $I_{3R}$ are broken by 
$\langle \chi^0 \rangle$, the combination 
\begin{equation}
I_{3R}+S = {D \over 3} 
\end{equation}
is unbroken.  
Although this idea was used previously~\cite{klm09,klm10}, the important 
observation here is that $I_{3R} + S$ coincides with the usual definition 
of $B - L$ for the known quarks and leptons, but takes on different values 
for the other particles.  Hence $D/3$ may be defined as generalized $B-L$ 
and functions as a global dark $U(1)$ symmetry.  Now
\begin{equation}
R_D = (-1)^{3B-3L+2j}
\end{equation}
so that it is identical to the usual definition of $R$ parity in supersymmetry 
for the SM particles.  Here the odd $R_D$ particles are the $h,n,x,y$ 
fermions, $(\eta^0,\eta^-),\lambda^0$ scalars, and $W_R^\pm$ vector bosons.
Note that leptonic color $SU(2)_l$ confines the $x,y$ hemions to bosons 
which must then have even $R_D$. 

To verify that generalized $B-L$ is indeed a global dark $U(1)$ symmetry of 
our model, consider the $SU(2)_R$ gauge bosons $(W_R^+,W_R^0,W_R^-)$ which 
has $S=0$.  Hence they have $I_{3R}+S$ values $(1,0,-1)$.  This is expected 
because $W_R^+$ takes $h_R$ to $u_R$ and $l_R$ to $n_R$.  Consider next 
the Yukawa terms allowed by the gauge symmetry and $S$, i.e.
\begin{eqnarray}
&& \bar{d}_R (u_L \phi_1^- - d_L \phi_1^0), ~ \bar{u}_R (u_L \phi_2^0 - 
d_L \phi_2^+) + \bar{h}_R (-u_L \eta^- + d_L \eta^0), ~ (\chi^+ \bar{u}_R 
- \chi^0 \bar{h}_R)h_L, \\ 
&& (\phi_1^0 \bar{\nu}_L + \phi_1^- \bar{l}_L)\nu_R, ~ \bar{\nu}_L (n_R 
\eta^0 + l_R \phi_2^+) + \bar{l}_L (n_R \eta^- + l_R \phi_2^0), ~ 
\bar{n}_L (n_R \chi^0 - l_R \chi^+), \\ 
&& \bar{z}_L (z_R \chi^0 - y_R \chi^+), ~ \bar{x}_R (\bar{z}_R \chi^+ 
+ \bar{y}_R \chi^0), ~~~ \bar{d}_R h_L \lambda^0, ~ \bar{n}_L \nu_R 
\lambda^0, ~~ \bar{z}_L x_R \lambda^0, ~~ z_R y_R \bar{\lambda}^0,
\end{eqnarray}
and the scalar trilinear terms
\begin{equation}
\phi_1^- (\eta^0 \chi^+ + \phi_2^+ \chi^0) - \phi_1^0 (\eta^- \chi^+ + 
\phi_2^0 \chi^0), ~~ \lambda^0(\eta^0 \phi_2^0 - \eta^- \phi_2^+).
\end{equation}
It is easily confirmed from the above that $I_{3R}+S$ is not broken by 
$\langle \phi^0_{1,2} \rangle$ and $\langle \chi^0 \rangle$.  Note that in 
the familar case of $SU(5)$ grand unification, neither $B$ nor $L$ is part 
of $SU(5)$ but both exist as low-energy conserved quantities. Here,  
$B$ and $L$ are again not part of $[SU(3)]^4$ separately, but 
a generalized $B-L$ emerges, and remains unbroken
to be naturally interpreted as a global dark symmetry.

\section{Gauge Sector}
Let
\begin{equation}
\langle \phi^0_1 \rangle = v_1, ~~~ \langle \phi^0_2 \rangle = v_2, ~~~ 
\langle \chi^0 \rangle = v_R,  
\end{equation}
then the $SU(3)_q \times SU(2)_l \times SU(2)_L \times SU(2)_R \times U(1)_X$ 
gauge symmetry is broken to $SU(3)_q \times SU(2)_l \times U(1)_{em}$ with 
a residual global $I_{3R}+S$ as the dark symmetry, as explained previously. 

Consider now the masses of the gauge bosons.  The charged ones, $W_L^\pm$ 
and $W_R^\pm$, do not mix because the latter have dark charge $\pm 1$.  
Their masses are given by
\begin{equation}
M_{W_L}^2 = {1 \over 2} g_L^2 (v_1^2 + v_2^2), ~~~ 
M_{W_R}^2 = {1 \over 2} g_R^2 (v_R^2 + v_2^2).
\end{equation}
Since $Q = I_{3L} + I_{3R} - X$, the photon is given by
\begin{equation}
A = {e \over g_L} W_{3L} + {e \over g_R} W_{3R} + {e \over g_X} Z_X,
\end{equation}
where $e^{-2} = g_L^{-2} + g_R^{-2} + g_X^{-2}$.  Let
\begin{eqnarray}
Z &=& (g_L^2 + g_Y^2)^{-1/2} \left( g_L W_{3L} - {g_Y^2 \over g_R} W_{3R} 
- {g_Y^2 \over g_X} Z_X \right), \\ 
Z' &=& (g_R^2 + g_X^2)^{-1/2} ( g_R W_{3R} - g_X Z_X),
\end{eqnarray}
where $g_Y^{-2} = g_R^{-2} + g_X^{-2}$, 
then the $2 \times 2$ mass-squared matrix spanning $(Z,Z')$ is given by
\begin{eqnarray}
{1 \over 2}\pmatrix{(g_L^2 + g_Y^2) (v_1^2 + v_2^2) & 
(\sqrt{g_L^2 + g_Y^2}/\sqrt{g_R^2 + g_X^2}) (g_X^2 v_1^2 - g_R^2 v_2^2) \cr  
(\sqrt{g_L^2 + g_Y^2}/\sqrt{g_R^2 + g_X^2}) (g_X^2 v_1^2 - g_R^2 v_2^2) & 
(g_R^2 + g_X^2) v_R^2 + (g_X^4 v_1^2 + g_R^4 v_2^2)/(g_R^2 + g_X^2)}. 
\end{eqnarray}
Their neutral-current interactions are given by
\begin{eqnarray}
{\cal L}_{NC} &=& e A_\mu j^\mu_Q + g_Z Z_\mu (j^\mu_{3L} - \sin^2 \theta_W 
j^\mu_{em}) + (g_R^2 + g_X^2)^{-1/2} Z'_\mu (g_R^2 j^\mu_{3R} + g_X^2 j^\mu_X),
\end{eqnarray}
where $g_Z^2 = g_L^2 + g_Y^2$ and $\sin^2 \theta_W = g_Y^2/g_Z^2$.
Since $Z-Z'$ mixing is constrained by experiment to be less than 
$10^{-4}$ or so, we assume $(g_X^2 v_1^2 - g_R^2 v_2^2)/v_R^2$ to be 
negligible.

The new gauge boson $Z'$ may be produced at the Large Hadron Collider (LHC) 
through their couplings to $u$ and $d$ quarks, and decay to charged leptons 
($e^-e^+$ and $\mu^-\mu^+$).  Hence current search limits for a $Z'$ boson are 
applicable.  Using $\alpha_R(M_R)=0.0290$ and $\alpha_X(M_R)=0.0163$, the 
$c_{u,d}$ coefficients~\cite{atlas14,cms15} used in the data analysis 
for our model are
\begin{equation}
c_u = (g_{uL}^2 + g_{uR}^2) B = 0.04~B, ~~~ 
c_d = (g_{dL}^2 + g_{dR}^2) B = 0.01~B,
\end{equation}
where $B$ is the branching fraction of $Z'$ to $e^-e^+$ and $\mu^-\mu^+$.
Assuming that $Z'$ decays to all the particles listed in Table 1, except 
for the scalars which become the longitudinal components of the various 
gauge bosons, we find $B = 0.044$.  Based on the 2016 LHC 13 TeV 
data set~\cite{atlas17}, this translates to a bound of about 3 to 4 TeV 
on the $Z'$ mass.

\section{Scalar Sector}
Consider the most general scalar potential consisting of 
$\Phi_L = (\phi_1^0,\phi_1^-)$, $\chi_R = (\chi^+,\chi^0)$, $\lambda^0$, and 
\begin{equation}
\eta = \pmatrix{\eta^0 & \phi_2^+ \cr \eta^- & \phi_2^0}, ~~~ 
\tilde{\eta} = \sigma_2 \eta^* \sigma_2 = \pmatrix{\bar{\phi}_2^0 & 
-\eta^+ \cr -\phi_2^- & \bar{\eta}^0},
\end{equation}
then
\begin{eqnarray}
V &=& -\mu^2_L \Phi_L^\dagger \Phi_L - \mu^2_R \chi_R^\dagger \chi_R - 
\mu^2_\eta Tr(\eta^\dagger \eta) - \mu^2_\lambda \bar{\lambda} \lambda
+ [\mu_1 \Phi_L^\dagger \eta \chi_R + \mu_2 \lambda det(\eta) + H.c.] 
\nonumber \\ 
&+& {1 \over 2} f_L (\Phi_L^\dagger \Phi_L)^2 + {1 \over 2} f_R 
(\chi_R^\dagger \chi_R)^2 + {1 \over 2} f_\lambda (\bar{\lambda} \lambda)^2 
+ {1 \over 2} f_\eta [Tr(\eta^\dagger \eta)]^2 + {1 \over 2} f'_\eta 
Tr(\eta^\dagger \eta \eta^\dagger \eta) \nonumber \\ 
&+& f_{LR} (\Phi_L^\dagger \Phi_L)(\chi_R^\dagger \chi_R) + f_{L\lambda} 
(\Phi_L^\dagger \Phi_L) (\bar{\lambda} \lambda) + f_{R\lambda} (\chi_R^\dagger 
\chi_R) (\bar{\lambda} \lambda) + f_{\lambda \eta} (\bar{\lambda} \lambda) 
Tr(\eta^\dagger \eta) \nonumber \\ 
&+& f_{L\eta} \Phi_L^\dagger \eta \eta^\dagger \Phi_L + f'_{L\eta} 
\Phi_L^\dagger \tilde{\eta} \tilde{\eta}^\dagger \Phi_L + f_{R\eta} 
\chi_R^\dagger \eta^\dagger \eta \chi_R + f'_{R\eta} 
\chi_R^\dagger \tilde{\eta}^\dagger \tilde{\eta} \chi_R.
\end{eqnarray}
Note that
\begin{eqnarray}
2 |det(\eta)|^2 &=& [Tr(\eta^\dagger \eta)]^2 - Tr(\eta^\dagger \eta \eta^\dagger 
\eta), \\ 
(\Phi_L^\dagger \Phi_L) Tr(\eta^\dagger \eta) &=& \Phi_L^\dagger \eta \eta^\dagger 
\Phi_L + \Phi_L^\dagger \tilde{\eta} \tilde{\eta}^\dagger \Phi_L, \\ 
(\chi_R^\dagger \chi_R) Tr(\eta^\dagger \eta) &=& \chi_R^\dagger \eta^\dagger \eta 
\chi_R + \chi_R^\dagger \tilde{\eta}^\dagger \tilde{\eta} \chi_R.
\end{eqnarray} 
The minimum of $V$ satisfies the conditions
\begin{eqnarray}
\mu_L^2 &=& f_L v_1^2 + f_{L\eta} v_2^2 + f_{LR} v_R^2 + \mu_1 v_2 v_R/v_1, \\ 
\mu_\eta^2 &=& (f_\eta + f'_\eta) v_2^2 + f_{L\eta} v_1^2 + 
f_{R\eta} v_R^2 + \mu_1 v_1 v_R/v_2, \\ 
\mu_R^2 &=& f_R v_R^2 + f_{LR} v_1^2 + f_{R\eta} v_2^2 + \mu_1 v_1 v_2/v_R.
\end{eqnarray}
The $3 \times 3$ mass-squared matrix spanning 
$\sqrt{2}Im(\phi_1^0,\phi_2^0,\chi^0)$ is then given by
\begin{equation}
{\cal M}^2_I = \mu_1 \pmatrix{-v_2v_R/v_1 & v_R & v_2 \cr v_R & -v_1v_R/v_2 
& v_1 \cr v_2 & v_1 & -v_1v_2/v_R}.
\end{equation}
and that spanning $\sqrt{2}Re(\phi_1^0,\phi_2^0,\chi^0)$ is 
\begin{equation}
{\cal M}^2_R = {\cal M}^2_I + 2\pmatrix{ f_L v_1^2 & f_{L\eta}v_1v_2 &  
f_{LR}v_1 v_R \cr f_{L\eta}v_1v_2 
& (f_\eta + f'_\eta)v_2^2 & f_{R\eta}v_2v_R \cr  f_{LR} v_1 v_R &  
f_{R\eta} v_2 v_R &  f_R v_R^2}. 
\end{equation}
Hence there are two zero eigenvalues in ${\cal M}^2_I$ with one nonzero 
eigenvalue $-\mu_1[v_1 v_2/v_R + v_R (v_1^2+v_2^2)/v_1v_2]$ corresponding 
to the eigenstate 
$(-v_1^{-1},v_2^{-1},v_R^{-1})/\sqrt{v_1^{-2}+v_2^{-2}+v_R^{-2}}$. 
In ${\cal M}^2_R$, the linear combination 
$H = (v_1,v_2,0)/\sqrt{v_1^2+v_2^2}$, 
is the standard-model Higgs boson, with 
\begin{equation}
m_H^2 = 2[f_L v_1^4 + (f_\eta + f'_\eta) v_2^4 + 2 f_{L\eta} 
v_1^2 v_2^2]/(v_1^2 + v_2^2).
\end{equation}
The other two scalar bosons are much heavier, with suppressed mixing to $H$, 
which may all be assumed to be small enough to avoid the constraints from 
dark-matter direct-search experiments.

The dark scalars are $\lambda^0$, $\chi^\pm$, and $(\eta^0,\eta^-)$.  
Whereas $\chi^\pm$ become the longitudinal components of $W_R^\pm$, 
the other scalars have the interaction
\begin{equation}
\mu_2 \lambda^0 (\eta^0 \phi_2^0 - \eta^- \phi_2^+) + H.c.
\end{equation}
The $2 \times 2$ mass-squared matrix linking $(\lambda,\bar{\eta})$ to 
$(\bar{\lambda},\eta)$ is given by
\begin{equation}
{\cal M}^2_{\lambda-\eta} = \pmatrix{-\mu_\lambda^2 + f_{L\lambda} v_1^2 + 
f_{R\lambda} v_R^2 + f_{\lambda \eta} v_2^2 & \mu_2 v_2 \cr \mu_2 v_2 
& -\mu_\eta^2 + f_\eta v_2^2 + f'_{L \eta} v_1^2 + f'_{R \eta} v_R^2}.
\end{equation} 
We assume $\mu_2$ to be very small so that there is negligible mixing, 
with $\lambda^0$ as the lighter particle which is our dark-matter 
candidate.  Note of course that $\eta^0$ is not a suitable candidate 
because it has $Z^0$ interactions.

\section{Dark Matter Interactions}
Consider the scalar singlet $\lambda^0$ as our dark-matter candidate. 
Let its coupling with the SM Higgs boson be $f_{\lambda H} \sqrt{2} v_H$, 
then it has been shown~\cite{kmppz17-1} that for $m_\lambda = 150$ GeV,  
$f_{\lambda H} < 4.4 \times 10^{-4}$ from the most recent direct-search 
result~\cite{x1t17}.  With such a small coupling, the $\lambda^0$ 
annihilation cross section in the early Universe through the SM Higgs 
boson is much too small for $\lambda^0$ to have the correct observed 
relic abundance.  Hence a different process is required.

Consider then the Yukawa sector.  As noted in Eq.~(36), the interactions 
$f_x \lambda^0 \bar{z}_L x_R$ and $f_y \bar{\lambda}^0 z_R y_R$ exist.  
Now $x_R/y_R$ forms a Dirac hemion and has been assumed to be light 
in the previous analysis on the renormalization-group running of gauge 
couplings.  For convenience, the outgoing $y_R$ may be redefined as 
incoming $x_L$.  Let $m_\lambda > m_x$, then 
$\lambda^0 \bar{\lambda}^0 \to x \bar{x}$ through $z$ 
exchange is possible as shown in Fig.~2. 
\begin{figure}[htb]
\vspace*{-5cm}
\hspace*{-3cm}
\includegraphics[scale=1]{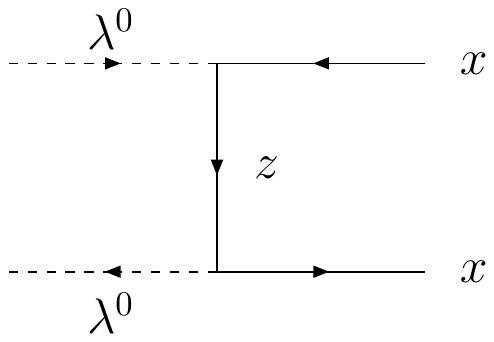}
\vspace*{-21.5cm}
\caption{Dark scalar annihilation to hemions.}
\end{figure}
\begin{figure}
\end{figure}
Let $f_y = f^*_x$ so that the $\lambda^0 \bar{z} x$ interaction is 
purely scalar.  The cross section $\times$ relative velocity is then given by
\begin{equation}
\sigma v_{rel} = {f_x^4 \over 4 \pi} \left( 1 - {m_x^2 \over m_\lambda^2} 
\right)^{3/2} {(m_z + m_x)^2 \over (m_z^2 + m_\lambda^2 - m_x^2)^2}.
\end{equation}
As an example, let $m_\lambda = 150$ GeV, $m_x = 100$ GeV, and $m_z = 600$ 
GeV, then $\sigma v_{rel} = 1$ pb is obtained for $f_x = 0.385$. 
The $x \bar{x}$ final states remain in thermal equilibrium through  
the photon, with their confined bound states (which are bosons with 
even $R_D$) decaying to SM particles as described in a following section. 

\section{Leptonic Color in the Early Universe}
As discussed in our earlier paper~\cite{kmppz17}, the $SU(2)_l$ massless 
stickons 
($\zeta$) play a role in the early Universe.  The important difference 
is that $\alpha_l(M_Z)$ is bigger here than in the Babu-Ma-Willenbrock 
(BMW) model~\cite{bmw04}, i.e. 0.065 
versus 0.047.  Hence the leptonic color confinement scale is about 
0.4 MeV instead of 4 keV.
At temperatures above the electroweak symmetry scale, the hemions are 
active and the stickons are in thermal equilibrium with the 
standard-model particles.  Below the hemion mass scale, the stickon 
interacts with photons through $\zeta \zeta \to \gamma \gamma$ 
scattering with a cross section
\begin{equation}
\sigma \sim {\alpha^2 \alpha_l^2 T^6 \over 64 m^8},
\end{equation}
where $m$ is the mass of the one light $x_R y_R$ hemion of this model. 
The decoupling temperature of $\zeta$ is then obtained by matching the 
Hubble expansion rate
\begin{equation}
H = \sqrt {(8 \pi/3) G_N (\pi^2/30) g_* T^4}
\end{equation}
to $[6\zeta(3)/\pi^2] T^3 \langle \sigma v \rangle$.  Hence 
\begin{equation}
T^{14} \sim {2^{12} \over 3^4} \left( {\pi^7 \over 5 [\zeta(3)]^2} \right) 
{G_N g_* m^{16} \over \alpha^4 \alpha_l^4}.
\end{equation}
For $m=100$ GeV and $g_* = 92.25$ which includes all particles with masses 
up to a few GeV,  $T \sim 9$ GeV.  Hence the contribution of 
stickons to the effective number of neutrinos at the time of big bang 
nucleosynthesis (BBN) is given by~\cite{jt13}
\begin{equation}
\Delta N_\nu = {8 \over 7} (3) \left( {10.75 \over 92.25} \right)^{4/3} 
= 0.195,
\end{equation}
compared to the value $0.50 \pm 0.23$ from a recent analysis~\cite{ns15}.  

As the Universe further cools below a few MeV, leptonic color goes through 
a phase transition and stickballs are formed.  However, they are not 
stable because they are allowed to mix with a scalar bound state of 
two hemions which would decay to two photons.  For a stickball $\omega$ of 
mass $m_\omega$, we assume this mixing to be $f_\omega m_\omega/m$, so that 
its decay rate is given by
\begin{equation}
\Gamma (\omega \to \gamma \gamma) = {\alpha^2 f_\omega^2 m_\omega^5 \over 
256 \pi^3 m^4}.
\end{equation}
Using $m_\omega = 1$ MeV as an example with $m=100$ GeV 
as before, its lifetime is estimated to be $1.0 \times 10^{7}s$ 
for $f_\omega=1$.  This means that it disappears long before 
the time of photon decoupling, so the $SU(2)_l$ sector contributes 
no additional relativistic degrees of freedom. Hence $N_{eff}$ remains 
the same as in the SM, i.e. 3.046, coming only from neutrinos. 
This agrees with the PLANCK measurement~\cite{planck16} of the 
cosmic microwave background (CMB), i.e. 
\begin{equation}
N_{eff} = 3.15 \pm 0.23.
\end{equation}

\section{Leptonic Color at Future $e^-e^+$ Colliders}
Unlike quarks, all hemions are heavy.  Hence the lightest bound state is 
likely to be at least 200 GeV.  Its cross section through electroweak 
production at the LHC is probably too small for it to be discovered. 
On the other hand, in analogy to the observations of $J/\psi$ and $\Upsilon$ 
at $e^-e^+$ colliders of the last century, the resonance production of 
the corresponding neutral vector bound states (hemionia) of these hemions 
is expected at a future $e^-e^+$ collider (ILC, CEPC, FCC-ee) 
with sufficient 
reach in total center-of-mass energy.  Their decays will be distinguishable 
from heavy quarkonia (such as toponia) experimentally.

As discussed in Ref.~\cite{kmppz17}, the formation of hemion bound states 
is analogous to that of 
QCD.  Instead of one-gluon exchange, the Coulomb potential binding a 
hemion-antihemion pair comes from one-stickon exchange.  The difference is 
just the change in an SU(3) color factor of 4/3 to an SU(2) color factor 
of 3/4.  The Bohr radius is then $a_0 = [(3/8) \bar{\alpha}_l m]^{-1}$, 
and the effective $\bar{\alpha}_l$ is defined by
\begin{equation} 
\bar {\alpha}_l = \alpha_l (a_0^{-1}).
\end{equation}
Using $\alpha_l (M_Z) = 0.065$ with $m=100$ GeV, 
we obtain $\bar{\alpha}_l = 0.087$ and $a_0^{-1} = 3.26$ GeV.  Consider 
the lowest-energy vector bound state $\Omega$ of the lightest hemion 
of mass $m=100$ GeV.  In analogy to the hydrogen atom, its binding energy 
is given by
\begin{equation}
E_b = {1 \over 4} \left( {3 \over 4} \right)^2 \bar{\alpha}_l^2 m 
= 106~{\rm MeV},
\end{equation}
and its wavefunction at the origin is
\begin{equation}
|\psi(0)|^2 = {1 \over \pi a_0^3} = 11.03~{\rm GeV}^3.
\end{equation}
Since $\Omega$ will appear as a narrow resonance at a future $e^-e^+$ 
collider, its observation depends on the integrated cross section over 
the energy range $\sqrt{s}$ around $m_\Omega$:
\begin{equation}
\int d \sqrt{s} ~\sigma (e^- e^+ \to \Omega \to X) = {6 \pi^2 \over 
m_\Omega^2} {\Gamma_{ee} \Gamma_X \over \Gamma_{tot}},
\end{equation}
where $\Gamma_{tot}$ is the total decay width of $\Omega$, and $\Gamma_{ee}$, 
$\Gamma_X$ are the respective partial widths.

Since $\Omega$ is a vector meson, it couples to both the photon and $Z$ boson 
through its constituent hemions.  Hence it will decay to $W^- W^+$, 
$q \bar{q}$, $l^- l^+$, and $\nu \bar{\nu}$.  Using
\begin{equation}
\langle 0 | \bar{x} \gamma^\mu x | \Omega \rangle = \epsilon_\Omega^\mu 
\sqrt{8 m_\Omega} |\psi(0)|,
\end{equation}
the $\Omega \to e^- e^+$ decay rate is given by
\begin{equation}
\Gamma (\Omega \to \gamma,Z \to e^- e^+) = {2 m_\Omega^2 \over 3 \pi} 
(|C_V|^2 + |C_A|^2) |\psi(0)|^2,
\end{equation}
where
\begin{eqnarray}
C_V &=& {e^2 (1/2) (-1) \over m_\Omega^2} + {g_Z^2 (-\sin^2 \theta_W/4)  
[(-1+4\sin^2 \theta_W)/4] \over m_\Omega^2 - M_Z^2}, \\ 
C_A &=& {g_Z^2 (-\sin^2 \theta_W/4)  (1/4) \over m_\Omega^2 - M_Z^2}.
\end{eqnarray}
In the above, $\Omega$ is composed of the singlet hemions $x_R$ and 
$y_R$ with invariant mass term $x_{1R} y_{2R} - x_{2R} y_{1R}$.  
The $(x_L,y_L)$ option, considered in the BMW model, is not available 
here because they are superheavy from the breaking of $SU(3)_L$ at $M_U$. 
Here $\Gamma_{ee} = 139$ eV.  Similar 
expressions hold for the other fermions of the SM.

For $\Omega \to W^- W^+$, the triple $\gamma W^- W^+$ and $Z W^- W^+$ vertices 
have the same structure.  The decay rate is calculated to be
\begin{equation}
\Gamma (\Omega \to \gamma,Z \to W^- W^+) = {m_\Omega^2 (1-r)^{3/2} \over 
6 \pi r^2} \left( 4 + 20r + 3 r^2 \right) 
C_W^2 |\psi(0)|^2,
\end{equation}
where $r = 4M_W^2/m_\Omega^2$ and 
\begin{equation}
C_W = {e^2 (1/2) \over m_\Omega^2} + {g_Z^2 (-\sin^2 \theta_W/4) \over 
m_\Omega^2 - M_Z^2}.
\end{equation}
Because of the accidental cancellation of the two terms in 
the above, $C_W$ turns out to be very small.  Hence $\Gamma_{WW} = 10$ eV.  
For $\Omega \to ZZ$, there is only the $t-$channel contribution, i.e.
\begin{equation}
\Gamma(\Omega \to ZZ) = {m_\Omega^2 (1-r_Z)^{5/2} \over 3 \pi r_Z} D_Z^2 
|\psi(0)|^2,
\end{equation}
where $r_Z = 4M_Z^2/m_\Omega^2$ and $D_Z = g_Z^2 \sin^4 \theta_W /
4(m_\Omega^2-2m_Z^2)$.   Hence $\Gamma_{ZZ}$ is negligible.
The $\Omega$ decay to two stickons is forbidden by charge conjugation. 
Its decay to three stickons is analogous to that of quarkonium to three 
gluons.  Whereas the latter forms a singlet which is symmetric in $SU(3)_C$, 
the former forms a singlet which is antisymmetric in $SU(2)_l$.  However, 
the two amplitudes are identical because the latter is symmetrized with 
respect to the exchange of the three gluons and the former is antisymmetrized 
with respect to the exchange of the three stickons.  Taking into account the 
different color factors of $SU(2)_l$ versus $SU(3)_C$, the decay rate of 
$\Omega$ to three stickons and to two stickons plus a photon are
\begin{eqnarray}
\Gamma (\Omega \to \zeta \zeta \zeta) &=& {16 \over 27} (\pi^2-9) {\alpha_l^3 
\over m_\Omega^2} |\psi(0)|^2, \\ 
\Gamma (\Omega \to \gamma \zeta \zeta) &=& {8 \over 9} (\pi^2-9) {\alpha 
\alpha_l^2 \over m_\Omega^2} |\psi(0)|^2. 
\end{eqnarray}
Hence $\Gamma_{\zeta \zeta \zeta} = 39$ eV and $\Gamma_{\gamma \zeta \zeta} = 7$ eV. 
The integrated cross section for $X = \mu^- \mu^+$ is then 
$1.2 \times 10^{-32}$ cm$^2$-keV.  For comparison, this number is 
$7.9 \times 10^{-30}$ cm$^2$-keV for the $\Upsilon(1S)$.  At a high-luminosity 
$e^- e^+$ collider, it should be feasible to make this observation.  
Table 2 summarizes all the partial decay widths.

\begin{table}[htb]
\caption{Partial decay widths of the hemionium $\Omega$.}
\begin{center}
\begin{tabular}{|c|c|}
\hline
Channel & Width \\
\hline
$\sum \nu \bar{\nu}$ & 36 eV  \\
\hline
$e^-e^+,\mu^-\mu^+,\tau^- \tau^+$ & 0.4 keV \\
\hline
$u \bar{u}, c \bar{c}$ & 0.3 keV \\
\hline
$d \bar{d}, s \bar{s}, b \bar{b}$ & 0.1 keV \\
\hline
$W^- W^+$ & 10 eV \\
$Z Z$ & $< 0.1$ eV \\
\hline
$\zeta \zeta \zeta$ & 39 eV \\ 
$\zeta \zeta \gamma$ & 7 eV \\
\hline
sum & 0.9 keV \\
\hline
\end{tabular}
\end{center}
\end{table}

There are important differences between QCD and QHD (quantum hemiodynamics). 
In the former, because of the existence of light $u$ and $d$ quarks, it is 
easy to pop up $u \bar{u}$ and $d \bar{d}$ pairs from the QCD vacuum.  
Hence the production of open charm in an $e^- e^+$ collider is described 
well by the fundamental process $e^- e^+ \to c \bar{c}$.  In the latter, 
there are no light hemions.  Instead it is easy to pop up the light 
stickballs from the QHD vacuum.  As a result, just above the threshold 
of making the $\Omega$ resonance, the many-body production of $\Omega$ + 
stickballs becomes possible.  This cross section is presumably also well 
described by the fundamental process $e^- e^+ \to x \bar{x}$, i.e.
\begin{eqnarray}
\sigma (e^- e^+ \to x \bar{x}) &=& {2 \pi \alpha^2 \over 3} 
\sqrt{1-{4m^2 \over s}}\left[ 
{(s + 2m^2) \over s^2} + {x_W^2 \over 2 (1-x_W)^2} {(s-m^2) \over (s-m_Z^2)^2} 
\right. 
\nonumber \\ &+& \left. {x_W \over (1-x_W)} {(s-m^2) \over s(s-m_Z^2)} - 
{(1-4x_W) \over 4(1-x_W)} {m^2 \over s(s-m_Z^2)} \right],
\end{eqnarray}
where $x_W = \sin^2 \theta_W$ and $s = 4E^2$ is the square of the 
center-of-mass energy.  
Using $m=100$ GeV and $s = (250~{\rm GeV})^2$ as an example, we find 
this cross section to be 0.79 pb.

In QCD, there are $q \bar{q}$ bound states which are bosons, and $qqq$ 
bound states which are fermions.  In QHD, there are only bound-state 
bosons, because the confining symmetry is $SU(2)_l$.  Also, unlike 
baryon (or quark) number in QCD, there is no such thing as hemion number 
in QHD, because $y$ is effectively $\bar{x}$.  This explains why there are 
no stable analog fermion in QHD such as the proton in QCD.

\section{Concluding Remarks}
Candidates for dark matter are often introduced in an {\it ad hoc} manner, 
because it is so easy to do.  There are thus numerous claimants to the 
title.  Is there a guiding principle?  One such is supersymmetry, where 
the superpartners of the SM particles naturally belong to a dark sector. 
Another possible guiding principle proposed recently  
is to look for a dark symmetry embedded as a gauge symmetry in a 
unifying extension of the SM, such as $[SU(3)]^N$.  In this paper, 
the alternative $[SU(3)]^4$ gauge model of leptonic color and dark 
matter is discussed in some detail.  The dark global $U(1)$ symmetry 
is identified as generalized $B-L$ and the dark parity is 
$R_D = (-1)^{3B-3L+2j}$. 
The dark sector contains fermions $(h,x,y,n)$, scalars 
$[(\eta^0,\eta^-),\lambda^0]$, and vector gauge bosons $W_R^\pm$, where 
$h$ is a dark quark of charge $-1/3$, $x,y$ are hemions of charge 
$\pm1/2$, and $n$ is a dark neutral fermion.  The dark matter of the Universe 
is presumably a neutral scalar dominated by the singlet $\lambda^0$. 

The absence of observations of new physics at the LHC is a possible 
indication that fundamental new physics may not be accessible using 
the strong interaction, i.e. quarks and gluons.  It is then natural to 
think about future $e^- e^+$ colliders.  But is there some fundamental 
issue of theoretical physics which may only reveal itself there? and not 
at hadron colliders?  The notion of leptonic color is such a possible 
answer.  Our alternative $[SU(3)]^4$ model allows for the existence of new 
half-charged fermions (hemions) under a confining $SU(2)_l$ leptonic 
color symmetry, with masses below the TeV scale.  It also predicts 
the $SU(2)_l$ confining scale to be 0.4 MeV, so that stickball bound states 
of the vector gauge stickons are formed.  These new particles have no 
QCD interactions, but hemions have electroweak couplings, so they 
are accessible in a future $e^- e^+$ collider, as described in this 
paper.

\section*{Acknowledgement}
This work was supported in part by the U.~S.~Department of Energy Grant 
No. DE-SC0008541.

\bibliographystyle{unsrt}

\begin{thebibliography}{99}
\bibitem{dgg84} A. De Rujula, H. Georgi, and S. L. Glashow, in 
{\it Fifth Workshop on Grand Unification}, edited by K. Kang, H. Fried, and 
P. Frampton (World Scientific, Singapore, 1984), p.~88.
\bibitem{bhp86} K. S. Babu, X.-G. He, and S. Pakvasa, Phys. Rev. {\bf D33}, 
763 (1986).
\bibitem{mmz04} E. Ma, M. Mondragon, and G. Zoupanos, JHEP {\bf 0412}, 
026 (2004).
\bibitem{bmw04} K. S. Babu, E. Ma, and S. Willenbrock, Phys. Rev. {\bf D69}, 
051301(R) (2004).
\bibitem{kmppz17} C. Kownacki, E. Ma, N. Pollard, O. Popov, and M. Zakeri, 
Phys. Lett. {\bf B769}, 267 (2017).
\bibitem{kmppz18} C. Kownacki, E. Ma, N. Pollard, O. Popov, and M. Zakeri, 
Phys. Lett. {\bf B777}, 121 (2018).
\bibitem{m05} E. Ma, Mod. Phys. Lett. {\bf A20}, 1953 (2005).
\bibitem{m04} E. Ma, Phys. Lett. {\bf B593}, 198 (2004).
\bibitem{m87} E. Ma, Phys. Rev. {\bf D36}, 274 (1987).
\bibitem{bhm87} K. S. Babu, X.-G. He, and E. Ma, Phys. Rev. {\bf D36}, 
878 (1987).
\bibitem{klm09} S. Khalil, H.-S. Lee, and E. Ma, Phys. Rev. {\bf D79}, 
041701(R) (2009).
\bibitem{klm10} S. Khalil, H.-S. Lee, and E. Ma, Phys. Rev. {\bf D81}, 
051702(R) (2010).
\bibitem{bmw14} S. Bhattacharya, E. Ma, and D. Wegman, Eur. Phys. J. 
{\bf C74}, 2902 (2014).
\bibitem{kmppz17-1} C. Kownacki, E. Ma, N. Pollard, O. Popov, and M. Zakeri, 
arXiv:1706.06501 [hep-ph].
\bibitem{fl90} R. Foot and H. Lew, Phys. Rev. {\bf D41}, 3502 (1990).
\bibitem{flv91} R. Foot, H. Lew, and R. R. Volkas, Phys. Rev. {\bf D44}, 1531 
(1991).
\bibitem{dhqvv18} P. V. Dong, T. Huong, F. Queiroz, J. W. F. Valle, and C. A. 
Vaquera-Araujo, arXiv:1710.06951.
\bibitem{b12} S. M. Barr, Phys. Rev. {\bf D85}, 013001 (2012).
\bibitem{m13} E. Ma, Phys. Rev. {\bf D88}, 117702 (2013).
\bibitem{m18} E. Ma, arXiv:1712.08994 [hep-ph].
\bibitem{m15} E. Ma, Phys. Rev. Lett. {\bf 115}, 011801 (2015).
\bibitem{atlas14} G. Aad {\it et al.} (ATLAS Collaboration), Phys. Rev. 
{\bf D90}, 052005 (2014).
\bibitem{cms15} S. Khachatryan {\it et al.} (CMS Collaboration), JHEP 
{\bf 1504}, 025 (2015).
\bibitem{atlas17} M. Aaboud {\it et al.} (ATLAS Collaboration), JHEP 
{\bf 1710}, 182 (2017).
\bibitem{x1t17} E. Aprile {\it et al.} (XENON Collaboration), Phys. Rev. 
Lett. {\bf 119}, 181301 (2017).
\bibitem{jt13} K. S. Jeong and F. Takahashi, Phys. Lett. {\bf B725}, 134 
(2013).
\bibitem{ns15} K. M. Nollett and G. Steigman, Phys. Rev. {\bf D91}, 083505 
(2015).
\bibitem{planck16} P. A. R. Ade {\it et al.} (PLANCK Collaboration), 
Astron.Astrophys. {\bf 594}, A13 (2016).
\end{thebibliography}

\end{document}